\documentclass[sigconf]{acmart}
\AtBeginDocument{%
  }

\copyrightyear{2025}
\acmYear{2025}
\setcopyright{rightsretained}
\acmConference[CSCW Companion '25]{Companion of the Computer-Supported Cooperative Work and Social Computing}{October 18--22, 2025}{Bergen, Norway}
\acmBooktitle{Companion of the Computer-Supported Cooperative Work and Social Computing (CSCW Companion '25), October 18--22, 2025, Bergen, Norway}\acmDOI{10.1145/3715070.3749237}
\acmISBN{979-8-4007-1480-1/2025/10}

\usepackage{graphicx}
\usepackage{subcaption}
\usepackage{float}
\usepackage{enumitem,multicol}
\setlist{nosep} % remove extra vertical space in lists
\newcommand{\FiveLikertLikely}{(Not at all likely -- Slightly likely -- Moderately likely -- Very likely -- Extremely likely)}
\newcommand{\FiveLikert}{(Not at all -- Slightly -- Moderately -- Very -- Extremely)}
\newcommand{\SevenEffort}{(1: Very little effort -- 7: A great deal of effort)}
\newcommand{\SevenAutoManual}{(1: Entirely automated -- 7: Entirely manual)}

\begin{document}

\title[Examining the Impact of Label Detail and Content Stakes on User
Perceptions \\ of AI-Generated Images on Social Media]{Examining the Impact of Label Detail and Content Stakes on User Perceptions of AI-Generated Images on Social Media}

\author{Jingruo Chen}
\email{jc3564@cornell.edu}
\affiliation{%
  \institution{Cornell University}
  \city{Ithaca}
  \state{NY}
  \country{USA}
}

\author{TungYen Wang}
\email{tw565@cornell.edu}
\affiliation{%
  \institution{Cornell University}
  \city{Ithaca}
  \state{NY}
  \country{USA}
}

\author{Marie Williams}
\email{mcw242@cornell.edu}
\affiliation{%
  \institution{Cornell University}
  \city{Ithaca}
  \state{NY}
  \country{USA}
}

\author{Natalia Jordan}
\email{naj46@cornell.edu}
\affiliation{%
  \institution{Cornell University}
  \city{Ithaca}
  \state{NY}
  \country{USA}
}

\author{Mingyi Shao}
\email{ms3737@cornell.edu}
\affiliation{%
  \institution{Cornell University}
  \city{Ithaca}
  \state{NY}
  \country{USA}
}

\author{Linda Zhang}
\email{lz324@cornell.edu}
\affiliation{%
  \institution{Cornell University}
  \city{Ithaca}
  \state{NY}
  \country{USA}
}

\author{Susan R. Fussell}
\email{sfussell@cornell.edu}
\affiliation{%
  \institution{Cornell University}
  \city{Ithaca}
  \state{NY}
  \country{USA}
}

\renewcommand{\shortauthors}{Jingruo Chen et al.}
%% No italics, no superscripts
%% Use footnote or author note to identify equal contribution and/or contact author info

\begin{abstract}
AI-generated images are increasingly prevalent on social media, raising concerns about trust and authenticity. This study investigates how different levels of label detail (basic, moderate, maximum) and content stakes (high vs. low) influence user engagement with and perceptions of AI-generated images through a within-subjects experimental study with 105 participants. Our findings reveal that increasing label detail enhances user perceptions of label transparency but does not affect user engagement. However, content stakes significantly impact user engagement and perceptions, with users demonstrating higher engagement and trust in low-stakes images. These results suggest that social media platforms can adopt detailed labels to improve transparency without compromising user engagement, offering insights for effective labeling strategies for AI-generated content.
\end{abstract}




\maketitle
\section{Background}
The rise of generative AI has transformed digital interactions, especially on social media, where AI-generated content (AIGC) ranges from photorealistic images to deepfakes, blurring the line between real and synthetic media \cite{fisher_moderating_2024, menczer_addressing_2023, yu_fake_2024}. This proliferation raises significant concerns about trust, credibility, and misinformation, making transparency and labeling essential for mitigating risks \cite{fisher_moderating_2024}. However, current labeling practices lack standardization, often failing to meet user expectations across different contexts \cite{burrus_unmasking_2024}.

Previous research underscores the impact of labeling on user perceptions. Rae showed that labels like "AI-assisted" or "human-created" affect user satisfaction and creator credibility but not perceived originality or trustworthiness \cite{rae_effects_2024}. Burrus et al. found that the effectiveness of the label varies with the stakes of the content: users are more critical of high-stakes content (e.g. news) than low-stakes scenarios (e.g. marketing)\cite{burrus_unmasking_2024}. Additionally, Feng et al. highlighted that detailed attributions, such as provenance information, lowered user trust, particularly for potentially deceptive media \cite{feng_examining_2023}. In addition to trust, perceptions of AI-generated content may be shaped by evolving expectations of creativity and labor in algorithmically mediated platforms. Prior work shows that social media norms, driven by visibility algorithms and audience feedback, shape how users and creators evaluate originality, authorship, and creative value \cite{de2024poets, simpson2023tiktok}. These expectations may also influence whether AI-generated content is seen as requiring meaningful effort. To examine this, our study asks participants to evaluate the perceived effort and manual work involved in creating each image.

Theoretical foundations like Explainable AI (XAI) and the Situation Awareness-based Agent Transparency (SAT) model \cite{chen_2014_situation} provide theoretical foundations for enhancing user understanding of AI-generated content. XAI emphasizes making AI processes understandable \cite{sanneman_situation_2022, borgo_towards_2018}, while SAT outlines three transparency levels: Goals and Actions, Reasoning Process, and Projections \cite{jiang_situation_2023}. These principles are helpful for labeling clarity, as user perceptions are shaped by the depth of transparency \cite{wittenberg_labeling_2024, tseng_investigating_2023}. In addition, user feedback is important for assessing labeling effectiveness. Clear, detailed labels can boost trust but may also trigger skepticism if seen as manipulative \cite{yaqub_effects_2020, shusas_trust_2024}. Our study extends this by collecting qualitative insights on user concerns and preferences for AI-generated content. 

Our study explores how labeling strategies affect user perceptions through a 2 (AI Image Content: Low Stake vs. High Stake) × 3 (Label Detail Level: Basic, Moderate, Maximum) within-subjects experiment with 105 participants. Participants viewed images and rated their engagement intention, perceptions of images, and label helpfulness. We also collected open-ended feedback on user concerns, preferences, and expectations for labeling. This study aims to address three research questions:

\begin{itemize}
    \item \textbf{RQ1:} How do different levels of label detail (basic, moderate, maximum) influence user engagement, perceptions of AI-generated images, and label helpfulness?
    \item \textbf{RQ2:} How do the stakes of AI-generated image content (high vs. low) impact user engagement and perceptions?
    \item \textbf{RQ3:} What are the user preferences regarding labeling for AI-generated images on social media?
\end{itemize}

Our findings show that more detailed labels for AIGC enhance label clarity, informativeness, and trustworthiness without reducing user engagement. Additionally, users engage more with and trust low-stakes content, while high-stakes content prompts greater scrutiny and skepticism. This suggests that platforms can enhance transparency with detailed labels without negatively impacting user engagement, supporting the responsible presentation of AI-generated content.

\section{Method}
We conducted an online, within-subjects experimental study on Qualtrics between November and December 2024, following approval from the Institutional Review Board (IRB). Participants were recruited through the SONA system, which offers research credit for university students.

\subsection{Participants}
A total of 105 participants completed the study. Most participants (92\%, $n = 96$) aged 18-22 (33\% male and 65\% female). Participants were primarily undergraduate students (96\%), with 62\% in STEM fields. Most (96\%) had encountered AI-generated content on social media, with exposure primarily through short videos (77\%), image posts (72\%), advertisements (42\%), and profile pictures (23\%). Participants' sentiments were mixed, with 57\% viewing AI’s prevalence negatively. Although 70\% had experimented with AI image generators primarily for entertainment or education, usage was infrequent, with 42\% using them 1-2 times.

\subsection{Materials}

\begin{figure*}
    \centering
    \includegraphics[width=1.8\columnwidth]{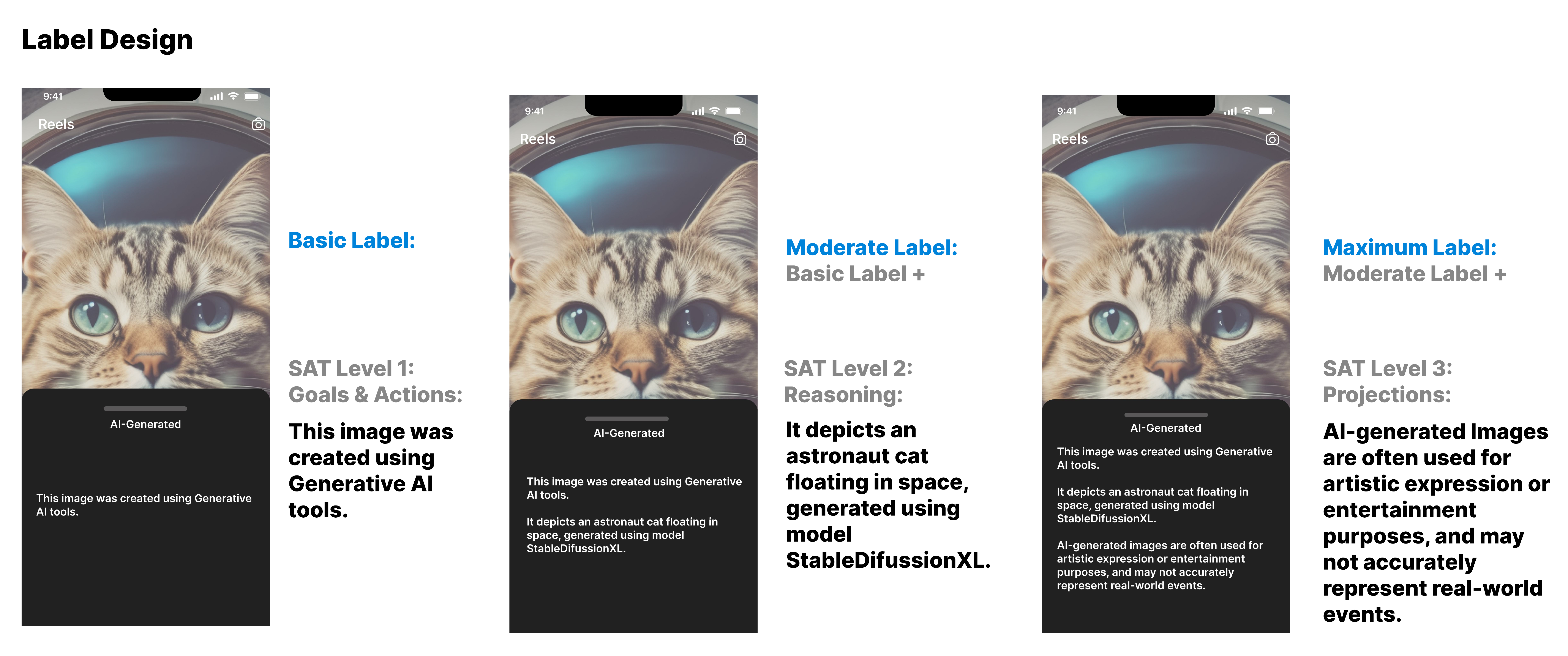}
    \caption{Three AI image label levels based on the SAT transparency model: Basic describes AI tool use (Level 1: Goals \& Actions), Moderate adds content reasoning and model used (Level 2: Reasoning), and Maximum includes broader implications (Level 3: Projections).}
    \label{fig:label_designs}
\end{figure*}

\subsubsection{Image Selection}
\label{Section:image selection}
Images were selected through a multi-step process. First, a pool of AI-generated images was created using DALL-E 3, Stable Diffusion, and MidJourney\footnote{DALL-E 3: https://openai.com/dall-e; Stable Diffusion: https://stability.ai/; MidJourney: https://www.midjourney.com/}, maintaining a consistent visual style. Second, images were categorized as low-stakes (e.g., scenery) or high-stakes (e.g., protests), reflecting benign versus socially or politically sensitive contexts known to influence user attention, credibility, and responses \cite{baron2025explainable, marmolejo2025factors}. Third, six researchers refined the pool for clarity, quality, and stake alignment, yielding three low-stakes (mountains, animals, portraits) and three high-stakes AI-generated images (protests, police, robot troops). Finally, each AI-generated image was paired with a matching real image from copyright-free sources (e.g., Shutterstock\footnote{Shutterstock: https://www.shutterstock.com/}), selected through the same consensus process to ensure content and style similarity. These real images served as fillers to mimic real-world social media but were not analyzed separately. All images are listed in Appendix \ref{appendix: survey questions} Figure \ref{fig: Image presented}.

\subsubsection{Label Design}
Labels for the AI-generated images were designed according to the Situation Awareness-Based Agent Transparency (SAT) model \cite{chen_2014_situation}. In the basic condition (SAT Level 1), labels indicated that the image was "AI-generated" without further elaboration. The moderate condition expanded this by providing a brief description of the image content and specifying the AI model used to generate it (SAT Level 1+2). The maximum condition further included a disclaimer stating that the image may not accurately represent real-world events (SAT Level 1+2+3). This progressive design enabled a systematic examination of how increasing label detail influenced user perceptions (Figure \ref{fig:label_designs}).

\subsection{Procedure}
Participants provided informed consent before viewing images in random order. Six AI-generated images were assigned to one of three labeling conditions (basic, moderate, maximum), with each condition applied to two images. Two real images from the curated pool (Section \ref{Section:image selection}) were shown without labels as fillers. After each image, participants rated engagement intentions (liking, sharing, commenting, seeking information), believability, manipulation, and creation effort. For AI-generated images, they also assessed label helpfulness, clarity, informativeness, and trustworthiness. Finally, participants completed a survey on demographics, familiarity with AI-generated content, and general attitudes toward AI, followed by four open-ended questions about their understanding of "AI-generated," related concerns, and label preferences.

\subsection{Measures}
Quantitative data were collected via Likert scales measuring engagement intentions (liking, sharing, commenting, seeking information; 1 = Not at all likely, 5 = Extremely likely), perceptions of believability and manipulation (1 = Not at all, 5 = Extremely), and creation effort/manual work (1 = Very little effort, 7 = A great deal). Label helpfulness was assessed through informativeness, clarity, and trustworthiness (1 = Not at all, 5 = Extremely), adapted from validated scales \cite{wittenberg_labeling_2024}. Qualitative data came from open-ended questions on participants’ understanding of “AI-generated,” related concerns, label design preferences, and traits considered trustworthy. Full survey questions appear in Appendix \ref{appendix: survey questions}.

\subsection{Data Analysis}
We used a mixed-methods approach to examine how label detail (basic, moderate, maximum) and image stakes (high vs. low) influenced user engagement (likes, shares, comments, seeking information), perceptions (believability, manipulation), and label helpfulness (clarity, informativeness, helpfulness, trustworthiness). Quantitative data were analyzed with a two-way mixed ANOVA, with label type and stake as independent variables and participant ID as a random effect; significant effects were followed by Tukey’s HSD tests. Qualitative data from open-ended responses underwent thematic analysis \cite{braun_using_2006}, with three researchers conducting iterative coding and theme refinement to ensure reliability.

\section{Findings}

\begin{table*}
\small
\renewcommand{\arraystretch}{1} % Adjust row height
\centering
\caption{ANOVA Results for User Engagement, Image Perceptions, and Label Clarity}
\begin{tabular}{l l r r r r r c}
\hline
\textbf{Dependent Variable} & \textbf{Source} & \textbf{SS} & \textbf{df} & \textbf{MS} & \textbf{F} & \textbf{p-value} & \textbf{Significance} \\
\hline
Share                      & Label Type          & 0.47   & 2  & 0.233 & 0.567  & 0.5674  & \\
                           & Stake               & 2.03   & 1  & 2.033 & 4.958  & 0.0264  & * \\
                           & Label Type × Stake  & 1.52   & 2  & 0.759 & 1.850  & 0.1583  & \\
                           & Error               & 207.85 & 507& 0.410 &        &         & \\
Like or Favourite          & Label Type          & 0.15   & 2  & 0.073 & 0.181  & 0.835   & \\
                           & Stake               & 25.15  & 1  & 25.15 & 62.286 & <0.001  & *** \\
                           & Label Type × Stake  & 1.03   & 2  & 0.514 & 1.273  & 0.281   & \\
                           & Error               & 204.68 & 507& 0.404 &        &         & \\
Comment                    & Label Type          & 0.04   & 2  & 0.019 & 0.097  & 0.907   & \\
                           & Stake               & 0.23   & 1  & 0.228 & 1.199  & 0.274   & \\
                           & Label Type × Stake  & 0.58   & 2  & 0.292 & 1.535  & 0.216   & \\
                           & Error               & 96.03  & 505& 0.190 &        &         & \\
Seek Additional Info       & Label Type          & 0.17   & 2  & 0.087 & 0.187  & 0.8299  & \\
                           & Stake               & 5.14   & 1  & 5.138 & 11.080 & <0.001  & *** \\
                           & Label Type × Stake  & 0.58   & 2  & 0.290 & 0.626  & 0.535   & \\
                           & Error               & 234.64 & 506& 0.464 &        &         & \\
Believable                 & Label Type          & 3.3    & 2  & 1.65  & 1.864  & 0.156   & \\
                           & Stake               & 32.2   & 1  & 32.21 & 36.456 & <0.001  & *** \\
                           & Label Type × Stake  & 0.4    & 2  & 0.18  & 0.207  & 0.813   & \\
                           & Error               & 448.8  & 508& 0.88  &        &         & \\
Manipulated                & Label Type          & 0.4    & 2  & 0.187 & 0.195  & 0.823   & \\
                           & Stake               & 14.9   & 1  & 14.91 & 15.592 & <0.001  & *** \\
                           & Label Type × Stake  & 3.6    & 2  & 1.822 & 1.905  & 0.150   & \\
                           & Error               & 484.9  & 507& 0.956 &        &         & \\
Effort                     & Label Type          & 0.5    & 2  & 0.251 & 0.288  & 0.750   & \\
                           & Stake               & 0.8    & 1  & 0.835 & 0.957  & 0.328   & \\
                           & Label Type × Stake  & 3.3    & 2  & 1.632 & 1.871  & 0.155   & \\
                           & Error               & 442.4  & 507& 0.872 &        &         & \\
Label Clarity              & Label Type          & 54.6   & 2  & 27.317& 23.907 & <0.001  & *** \\
                           & Stake               & 0.1    & 1  & 0.109 & 0.095  & 0.758   & \\
                           & Label Type × Stake  & 0.1    & 2  & 0.033 & 0.029  & 0.972   & \\
                           & Error               & 578.2  & 506& 1.143 &        &         & \\
\hline
\multicolumn{8}{l}{\textit{Signif. codes:  0 ‘***’ 0.001 ‘**’ 0.01 ‘*’ 0.05 ‘.’ 0.1 ‘ ’ 1}} \\
\end{tabular}
\end{table*}

\subsection{RQ1: Impact of Different Levels of Label Detail}
Maximum-detail labels consistently outperform both moderate and basic labels in label helpfulness. Maximum labels were \textbf{clearer} ($F(2, 506) = 23.907, p < 0.001$) than moderate (estimate = -0.583, SE = 0.168, $p = 0.0017$) and basic labels (estimate = -0.697, SE = 0.169, $p = 0.0001$). They were \textbf{more informative} ($F(2, 505) = 47.88, p < 0.001$) than moderate (estimate = -0.451, SE = 0.177, $p = 0.0298$) and basic (estimate = -1.147, SE = 0.178, $p < 0.001$). For \textbf{helpfulness} ($F(2, 507) = 36.689, p < 0.001$), maximum labels were better than basic (estimate = -0.980, SE = 0.178, $p < 0.001$) but not significantly different from moderate (estimate = -0.377, SE = 0.177, $p = 0.0832$). In \textbf{trustworthiness} ($F(2, 506) = 16.902, p < 0.001$), maximum labels outperformed basic (estimate = -0.5475, SE = 0.175, $p = 0.0054$) but were similar to moderate. Label types did not affect engagement (shares: $F(2, 506) = 0.567, p = 0.5674$; likes: $F(2, 506) = 0.181, p = 0.835$; comments: $F(2, 506) = 0.097, p = 0.907$; info seeking: $F(2, 506) = 0.187, p = 0.8299$) or image perceptions (believable: $F(2, 506) = 1.864, p = 0.156$; manipulated: $F(2, 506) = 0.195, p = 0.823$; effort: $F(2, 507) = 0.288, p = 0.750$; manual work: $F(2, 507) = 0.176, p = 0.839$).

\subsection{RQ2: Impact of Different Contents (High stake v.s. Low stake) of AI-generated Images}

User engagement and image believability varied significantly by image stakes. Low-stake images received more \textbf{likes} ($F(1, 507) = 62.286, p < 0.001$; estimate = 0.391, SE = 0.0704, $p < 0.001$), while high-stake images led to more \textbf{seeking of information} ($F(1, 506) = 11.080, p < 0.001$; estimate = -0.188, SE = 0.0714, $p = 0.0086$). Low-stake images appeared more \textbf{believable} ($F(1, 508) = 36.456, p < 0.001$; estimate = 0.447, SE = 0.0916, $p < 0.001$), whereas high-stake images seemed more \textbf{manipulated} ($F(1, 507) = 15.592, p < 0.001$; estimate = -0.306, SE = 0.0965, $p = 0.0016$). For shares, stakes had a significant main effect ($F(1, 507) = 4.958, p = 0.0264$), but post-hoc tests showed no significant difference (estimate = 0.105, SE = 0.0697, $p = 0.1309$). Stake did not significantly influence perceptions of label clarity ($F(1, 506) = 0.095, p = 0.758$), informativeness ($F(1, 506) = 0.288, p = 0.592$), helpfulness ($F(1, 507) = 0.001, p = 0.973$), trustworthiness ($F(1, 506) = 2.177, p = 0.141$), effort ($F(1, 507) = 0.957, p = 0.328$), or manual work ($F(1, 507) = 0.002, p = 0.967$).

\subsection{RQ3: User Preferences regarding labeling for AI-generated Images}

Users were mainly concerned about AI-generated images spreading fake news and deepfakes (61\%), with 33\% struggling to distinguish them from real content. Other concerns included reputational harm, exploitation, and explicit fakes (16\%), risks to vulnerable groups (10\%), and creativity erosion (7\%). Transparency was critical for trust: 40\% emphasized disclosing technical details (e.g., tools or models used). As one user noted, \textit{“If I can Google the model that a user used, I can better understand the creation of the image, which makes it less scary.”} Clear AI indicators were preferred by 36\%, while 24\% wanted clarity on whether an image was fully or partially AI-generated, saying, \textit{“Whether the image was entirely made by AI or AI only played a part in modifying it.”} Additionally, 15\% emphasized understanding the image’s intent (e.g., satire, education).

\section{Discussion}
Our findings demonstrate that while label type does not significantly impact user engagement, it strongly influences user perceptions. Maximum-detail labels consistently outperformed moderate and basic labels in terms of informativeness and trustworthiness, emphasizing transparency as crucial for user trust in AI \cite{molina2022ai}. However, this enhanced cognitive understanding did not translate into increased engagement, consistent with previous findings \cite{gamage2025labeling}. This suggests that while users cognitively recognize and value transparency, engagement may be more strongly driven by emotional relevance or social motivations than by informational clarity alone. Content stakes also influenced perceptions: low-stakes images were seen as more believable and less manipulated, reflecting greater skepticism toward AI-generated content in high-stakes contexts like news \cite{marmolejo2025factors, reuters2024ai}. This may stem from reduced perceived risk \cite{kalpokas2021synthetic}, increased familiarity \cite{horowitz2024adopting}, and lower emotional investment \cite{zhang2024decoding} in low-stakes content. Finally, the lack of significant effects on perceived effort suggests that transparency does not undermine perceptions of creator skill, consistent with studies indicating that transparency can enhance rather than undermine perceptions of expertise  \cite{montecchi2024perceived}.

\subsection{Design Implications}
Our findings provide actionable guidance for platforms and creators. First, platforms can use maximum-detail labels to enhance transparency without fearing reduced engagement, especially in critical contexts like news, education, or health. For example, platforms may provide specific information on the AI model used, the generation method, any human modifications involved, and a reality warning, conveying the nature of the content. Second, labeling strategies should adapt based on content stakes, as detailed labels may be more critical for high-stakes content. Third, creators can specify the creation process of their content without risking lower engagement or perceptions of effort.

\subsection{Limitations and Future Research}
This study has several limitations. Our participant pool primarily comprised undergraduate STEM students (96\%), who may have higher-than-average familiarity with AI technologies, potentially limiting the generalizability of our findings. Additionally, our study used a small, fixed set of six AI-generated images, which may have constrained variability in user responses. Moreover, the forced-exposure design does not fully reflect real-world conditions, where labels can be subtle or easily ignored. Prior research suggests that label effectiveness may decrease over time as users become habituated \cite{grady_nevertheless_2021}. Future studies should address these limitations by including diverse populations, testing labeling strategies in naturalistic settings, and exploring adaptive labeling approaches that can respond to user behavior over time. Future work should also examine how label design performs across different placements and saliency levels (e.g., captions, overlays, or metadata) in attention-limited contexts like mobile feeds. Expanding the scope of content to include emotionally charged or misleading images, such as satirical memes or synthetic political content, would further clarify how users interpret and act on AI transparency signals.

\bibliographystyle{ACM-Reference-Format}
\bibliography{GenAI_Label}


\begin{thebibliography}{27}

%%% the Web address.

\ifx \showCODEN    \undefined \def \showCODEN     #1{\unskip}     \fi
\ifx \showISBNx    \undefined \def \showISBNx     #1{\unskip}     \fi
\ifx \showISBNxiii \undefined \def \showISBNxiii  #1{\unskip}     \fi
\ifx \showISSN     \undefined \def \showISSN      #1{\unskip}     \fi
\ifx \showLCCN     \undefined \def \showLCCN      #1{\unskip}     \fi
\ifx \shownote     \undefined \def \shownote      #1{#1}          \fi
\ifx \showarticletitle \undefined \def \showarticletitle #1{#1}   \fi
\ifx \showURL      \undefined \def \showURL       {\relax}        \fi
% invisible to TeX
\providecommand\bibfield[2]{#2}
\providecommand\bibinfo[2]{#2}
\providecommand\natexlab[1]{#1}
\providecommand\showeprint[2][]{arXiv:#2}

\bibitem[Baron et~al\mbox{.}(2025)]%
        {baron2025explainable}
\bibfield{author}{\bibinfo{person}{Sam Baron}, \bibinfo{person}{Andrew~J Latham}, {and} \bibinfo{person}{Somogy Varga}.} \bibinfo{year}{2025}\natexlab{}.
\newblock \showarticletitle{Explainable AI and stakes in medicine: A user study}.
\newblock \bibinfo{journal}{\emph{Artificial Intelligence}}  \bibinfo{volume}{340} (\bibinfo{year}{2025}), \bibinfo{pages}{104282}.
\newblock


\bibitem[Borgo et~al\mbox{.}(2018)]%
        {borgo_towards_2018}
\bibfield{author}{\bibinfo{person}{Rita Borgo}, \bibinfo{person}{Michael Cashmore}, {and} \bibinfo{person}{Daniele Magazzeni}.} \bibinfo{year}{2018}\natexlab{}.
\newblock \bibinfo{title}{Towards Providing Explanations for {AI} Planner Decisions}.
\newblock
\href{https://doi.org/10.48550/arXiv.1810.06338}{doi:\nolinkurl{10.48550/arXiv.1810.06338}}
\showeprint[arxiv]{1810.06338}


\bibitem[Braun and Clarke(2006)]%
        {braun_using_2006}
\bibfield{author}{\bibinfo{person}{Virginia Braun} {and} \bibinfo{person}{Victoria Clarke}.} \bibinfo{year}{2006}\natexlab{}.
\newblock \showarticletitle{Using Thematic Analysis in Psychology}.
\newblock \bibinfo{journal}{\emph{Qualitative Research in Psychology}} \bibinfo{volume}{3}, \bibinfo{number}{2} (\bibinfo{year}{2006}), \bibinfo{pages}{77--101}.
\newblock
\showISSN{1478-0895}
\href{https://doi.org/10.1191/1478088706qp063oa}{doi:\nolinkurl{10.1191/1478088706qp063oa}}


\bibitem[Burrus et~al\mbox{.}(2024)]%
        {burrus_unmasking_2024}
\bibfield{author}{\bibinfo{person}{Olivia Burrus}, \bibinfo{person}{Amanda Curtis}, {and} \bibinfo{person}{Laura Herman}.} \bibinfo{year}{2024}\natexlab{}.
\newblock \showarticletitle{Unmasking {AI}: Informing Authenticity Decisions by Labeling {AI}-Generated Content}.
\newblock \bibinfo{journal}{\emph{Interactions}} \bibinfo{volume}{31}, \bibinfo{number}{4} (\bibinfo{year}{2024}), \bibinfo{pages}{38--42}.
\newblock
\showISSN{1072-5520}
\href{https://doi.org/10.1145/3665321}{doi:\nolinkurl{10.1145/3665321}}


\bibitem[Chen et~al\mbox{.}(2014)]%
        {chen_2014_situation}
\bibfield{author}{\bibinfo{person}{Jessie~Y Chen}, \bibinfo{person}{Katelyn Procci}, \bibinfo{person}{Michael Boyce}, \bibinfo{person}{Julia Wright}, \bibinfo{person}{Andre Garcia}, {and} \bibinfo{person}{Michael Barnes}.} \bibinfo{year}{2014}\natexlab{}.
\newblock \showarticletitle{Situation Awareness-Based Agent Transparency}.
\newblock \bibinfo{journal}{\emph{US Army Research Laboratory}} \bibinfo{number}{April} (\bibinfo{year}{2014}), \bibinfo{pages}{1--29}.
\newblock


\bibitem[De and Lu(2024)]%
        {de2024poets}
\bibfield{author}{\bibinfo{person}{Ankolika De} {and} \bibinfo{person}{Zhicong Lu}.} \bibinfo{year}{2024}\natexlab{}.
\newblock \showarticletitle{\#PoetsOfInstagram: Navigating The Practices And Challenges Of Novice Poets On Instagram}. In \bibinfo{booktitle}{\emph{Proceedings of the 2024 CHI Conference on Human Factors in Computing Systems}} (Honolulu, HI, USA) \emph{(\bibinfo{series}{CHI '24})}. \bibinfo{publisher}{Association for Computing Machinery}, \bibinfo{address}{New York, NY, USA}, Article \bibinfo{articleno}{162}, \bibinfo{numpages}{16}~pages.
\newblock
\showISBNx{9798400703300}
\href{https://doi.org/10.1145/3613904.3642173}{doi:\nolinkurl{10.1145/3613904.3642173}}


\bibitem[Feng et~al\mbox{.}(2023)]%
        {feng_examining_2023}
\bibfield{author}{\bibinfo{person}{K.~J.~Kevin Feng}, \bibinfo{person}{Nick Ritchie}, \bibinfo{person}{Pia Blumenthal}, \bibinfo{person}{Andy Parsons}, {and} \bibinfo{person}{Amy~X. Zhang}.} \bibinfo{year}{2023}\natexlab{}.
\newblock \showarticletitle{Examining the Impact of Provenance-Enabled Media on Trust and Accuracy Perceptions}.
\newblock \bibinfo{journal}{\emph{Proceedings of the ACM on Human-Computer Interaction}}  \bibinfo{volume}{7} (\bibinfo{year}{2023}), \bibinfo{pages}{270:1--270:42}.
\newblock
Issue CSCW2.
\href{https://doi.org/10.1145/3610061}{doi:\nolinkurl{10.1145/3610061}}


\bibitem[Fisher et~al\mbox{.}(2024)]%
        {fisher_moderating_2024}
\bibfield{author}{\bibinfo{person}{Sarah~A. Fisher}, \bibinfo{person}{Jeffrey~W. Howard}, {and} \bibinfo{person}{Beatriz Kira}.} \bibinfo{year}{2024}\natexlab{}.
\newblock \showarticletitle{Moderating Synthetic Content: The Challenge of Generative {AI}}.
\newblock \bibinfo{journal}{\emph{Philosophy \& Technology}} \bibinfo{volume}{37}, \bibinfo{number}{4} (\bibinfo{year}{2024}), \bibinfo{pages}{133}.
\newblock
\showISSN{2210-5441}
\href{https://doi.org/10.1007/s13347-024-00818-9}{doi:\nolinkurl{10.1007/s13347-024-00818-9}}


\bibitem[Gamage et~al\mbox{.}(2025)]%
        {gamage2025labeling}
\bibfield{author}{\bibinfo{person}{Dilrukshi Gamage}, \bibinfo{person}{Dilki Sewwandi}, \bibinfo{person}{Min Zhang}, {and} \bibinfo{person}{Arosha~K Bandara}.} \bibinfo{year}{2025}\natexlab{}.
\newblock \showarticletitle{Labeling Synthetic Content: User Perceptions of Label Designs for AI-Generated Content on Social Media}. In \bibinfo{booktitle}{\emph{Proceedings of the 2025 CHI Conference on Human Factors in Computing Systems}}. \bibinfo{pages}{1--29}.
\newblock


\bibitem[Grady et~al\mbox{.}(2021)]%
        {grady_nevertheless_2021}
\bibfield{author}{\bibinfo{person}{Rebecca~Hofstein Grady}, \bibinfo{person}{Peter~H. Ditto}, {and} \bibinfo{person}{Elizabeth~F. Loftus}.} \bibinfo{year}{2021}\natexlab{}.
\newblock \showarticletitle{Nevertheless, Partisanship Persisted: Fake News Warnings Help Briefly, but Bias Returns with Time}.
\newblock \bibinfo{journal}{\emph{Cognitive Research: Principles and Implications}} \bibinfo{volume}{6}, \bibinfo{number}{1} (\bibinfo{year}{2021}), \bibinfo{pages}{52}.
\newblock
\showISSN{2365-7464}
\href{https://doi.org/10.1186/s41235-021-00315-z}{doi:\nolinkurl{10.1186/s41235-021-00315-z}}


\bibitem[Horowitz et~al\mbox{.}(2024)]%
        {horowitz2024adopting}
\bibfield{author}{\bibinfo{person}{Michael~C Horowitz}, \bibinfo{person}{Lauren Kahn}, \bibinfo{person}{Julia Macdonald}, {and} \bibinfo{person}{Jacquelyn Schneider}.} \bibinfo{year}{2024}\natexlab{}.
\newblock \showarticletitle{Adopting AI: how familiarity breeds both trust and contempt}.
\newblock \bibinfo{journal}{\emph{AI \& society}} \bibinfo{volume}{39}, \bibinfo{number}{4} (\bibinfo{year}{2024}), \bibinfo{pages}{1721--1735}.
\newblock


\bibitem[Jiang et~al\mbox{.}(2023)]%
        {jiang_situation_2023}
\bibfield{author}{\bibinfo{person}{Jinglu Jiang}, \bibinfo{person}{Alexander~J. Karran}, \bibinfo{person}{Constantinos~K. Coursaris}, \bibinfo{person}{Pierre-Majorique Léger}, {and} \bibinfo{person}{Joerg Beringer}.} \bibinfo{year}{2023}\natexlab{}.
\newblock \showarticletitle{A Situation Awareness Perspective on Human-{AI} Interaction: Tensions and Opportunities}.
\newblock \bibinfo{journal}{\emph{International Journal of Human–Computer Interaction}} \bibinfo{volume}{39}, \bibinfo{number}{9} (\bibinfo{year}{2023}), \bibinfo{pages}{1789--1806}.
\newblock
\showISSN{1044-7318}
\href{https://doi.org/10.1080/10447318.2022.2093863}{doi:\nolinkurl{10.1080/10447318.2022.2093863}}


\bibitem[Kalpokas and Kalpokiene(2021)]%
        {kalpokas2021synthetic}
\bibfield{author}{\bibinfo{person}{Ignas Kalpokas} {and} \bibinfo{person}{Julija Kalpokiene}.} \bibinfo{year}{2021}\natexlab{}.
\newblock \showarticletitle{Synthetic media and information warfare: Assessing potential threats}.
\newblock \bibinfo{journal}{\emph{The Russian Federation in Global Knowledge Warfare: Influence Operations in Europe and Its Neighbourhood}} (\bibinfo{year}{2021}), \bibinfo{pages}{33--50}.
\newblock


\bibitem[Marmolejo-Ramos et~al\mbox{.}(2025)]%
        {marmolejo2025factors}
\bibfield{author}{\bibinfo{person}{Fernando Marmolejo-Ramos}, \bibinfo{person}{Rebecca Marrone}, \bibinfo{person}{Malgorzata Korolkiewicz}, \bibinfo{person}{Florence Gabriel}, \bibinfo{person}{George Siemens}, \bibinfo{person}{Srecko Joksimovic}, \bibinfo{person}{Yuki Yamada}, \bibinfo{person}{Yuki Mori}, \bibinfo{person}{Talal Rahwan}, \bibinfo{person}{Maria Sahakyan}, {et~al\mbox{.}}} \bibinfo{year}{2025}\natexlab{}.
\newblock \showarticletitle{Factors influencing trust in algorithmic decision-making: an indirect scenario-based experiment}.
\newblock \bibinfo{journal}{\emph{Frontiers in Artificial Intelligence}}  \bibinfo{volume}{7} (\bibinfo{year}{2025}), \bibinfo{pages}{1465605}.
\newblock


\bibitem[Menczer et~al\mbox{.}(2023)]%
        {menczer_addressing_2023}
\bibfield{author}{\bibinfo{person}{Filippo Menczer}, \bibinfo{person}{David Crandall}, \bibinfo{person}{Yong-Yeol Ahn}, {and} \bibinfo{person}{Apu Kapadia}.} \bibinfo{year}{2023}\natexlab{}.
\newblock \showarticletitle{Addressing the Harms of {AI}-Generated Inauthentic Content}.
\newblock \bibinfo{journal}{\emph{Nature Machine Intelligence}} \bibinfo{volume}{5}, \bibinfo{number}{7} (\bibinfo{year}{2023}), \bibinfo{pages}{679--680}.
\newblock
\showISSN{2522-5839}
\href{https://doi.org/10.1038/s42256-023-00690-w}{doi:\nolinkurl{10.1038/s42256-023-00690-w}}


\bibitem[Molina and Sundar(2022)]%
        {molina2022ai}
\bibfield{author}{\bibinfo{person}{Maria~D Molina} {and} \bibinfo{person}{S~Shyam Sundar}.} \bibinfo{year}{2022}\natexlab{}.
\newblock \showarticletitle{When AI moderates online content: effects of human collaboration and interactive transparency on user trust}.
\newblock \bibinfo{journal}{\emph{Journal of Computer-Mediated Communication}} \bibinfo{volume}{27}, \bibinfo{number}{4} (\bibinfo{year}{2022}), \bibinfo{pages}{zmac010}.
\newblock


\bibitem[Montecchi et~al\mbox{.}(2024)]%
        {montecchi2024perceived}
\bibfield{author}{\bibinfo{person}{Matteo Montecchi}, \bibinfo{person}{Kirk Plangger}, \bibinfo{person}{Douglas West}, {and} \bibinfo{person}{Ko de Ruyter}.} \bibinfo{year}{2024}\natexlab{}.
\newblock \showarticletitle{Perceived brand transparency: A conceptualization and measurement scale}.
\newblock \bibinfo{journal}{\emph{Psychology \& Marketing}} \bibinfo{volume}{41}, \bibinfo{number}{10} (\bibinfo{year}{2024}), \bibinfo{pages}{2274--2297}.
\newblock


\bibitem[Rae(2024)]%
        {rae_effects_2024}
\bibfield{author}{\bibinfo{person}{Irene Rae}.} \bibinfo{year}{2024}\natexlab{}.
\newblock \showarticletitle{The Effects of Perceived AI Use on Content Perceptions}. In \bibinfo{booktitle}{\emph{Proceedings of the 2024 CHI Conference on Human Factors in Computing Systems}} (Honolulu, HI, USA) \emph{(\bibinfo{series}{CHI '24})}. \bibinfo{publisher}{Association for Computing Machinery}, \bibinfo{address}{New York, NY, USA}, Article \bibinfo{articleno}{978}, \bibinfo{numpages}{14}~pages.
\newblock
\showISBNx{9798400703300}
\href{https://doi.org/10.1145/3613904.3642076}{doi:\nolinkurl{10.1145/3613904.3642076}}


\bibitem[{Reuters Institute for the Study of Journalism}(2024)]%
        {reuters2024ai}
\bibfield{author}{\bibinfo{person}{{Reuters Institute for the Study of Journalism}}.} \bibinfo{year}{2024}\natexlab{}.
\newblock \bibinfo{title}{Public Attitudes Towards the Use of AI in Journalism}.
\newblock
\urldef\tempurl%
\url{https://reutersinstitute.politics.ox.ac.uk/digital-news-report/2024/public-attitudes-towards-use-ai-and-journalism}
\showURL{%
\tempurl}


\bibitem[Sanneman and Shah(2022)]%
        {sanneman_situation_2022}
\bibfield{author}{\bibinfo{person}{Lindsay Sanneman} {and} \bibinfo{person}{Julie~A. Shah}.} \bibinfo{year}{2022}\natexlab{}.
\newblock \showarticletitle{The Situation Awareness Framework for Explainable {AI} ({SAFE}-{AI}) and Human Factors Considerations for {XAI} Systems}.
\newblock \bibinfo{journal}{\emph{International Journal of Human–Computer Interaction}} \bibinfo{volume}{38}, \bibinfo{number}{18} (\bibinfo{year}{2022}), \bibinfo{pages}{1772--1788}.
\newblock
\showISSN{1044-7318}
\href{https://doi.org/10.1080/10447318.2022.2081282}{doi:\nolinkurl{10.1080/10447318.2022.2081282}}


\bibitem[Shusas and Forte(2024)]%
        {shusas_trust_2024}
\bibfield{author}{\bibinfo{person}{Erica Shusas} {and} \bibinfo{person}{Andrea Forte}.} \bibinfo{year}{2024}\natexlab{}.
\newblock \showarticletitle{Trust and Transparency: An Exploratory Study on Emerging Adults' Interpretations of Credibility Indicators on Social Media Platforms}. In \bibinfo{booktitle}{\emph{Extended Abstracts of the CHI Conference on Human Factors in Computing Systems}} \emph{(\bibinfo{series}{CHI EA '24})}. \bibinfo{publisher}{Association for Computing Machinery}, \bibinfo{address}{New York, NY, USA}, \bibinfo{pages}{1--7}.
\newblock
\showISBNx{9798400703317}
\href{https://doi.org/10.1145/3613905.3650801}{doi:\nolinkurl{10.1145/3613905.3650801}}


\bibitem[Simpson and Semaan(2023)]%
        {simpson2023tiktok}
\bibfield{author}{\bibinfo{person}{Ellen Simpson} {and} \bibinfo{person}{Bryan Semaan}.} \bibinfo{year}{2023}\natexlab{}.
\newblock \showarticletitle{Rethinking Creative Labor: A Sociotechnical Examination of Creativity \& Creative Work on TikTok}. In \bibinfo{booktitle}{\emph{Proceedings of the 2023 CHI Conference on Human Factors in Computing Systems}} (Hamburg, Germany) \emph{(\bibinfo{series}{CHI '23})}. \bibinfo{publisher}{Association for Computing Machinery}, \bibinfo{address}{New York, NY, USA}, Article \bibinfo{articleno}{244}, \bibinfo{numpages}{16}~pages.
\newblock
\showISBNx{9781450394215}
\href{https://doi.org/10.1145/3544548.3580649}{doi:\nolinkurl{10.1145/3544548.3580649}}


\bibitem[Tseng and Yuan(2023)]%
        {tseng_investigating_2023}
\bibfield{author}{\bibinfo{person}{Yu-Chia Tseng} {and} \bibinfo{person}{Chien Wen~(Tina) Yuan}.} \bibinfo{year}{2023}\natexlab{}.
\newblock \showarticletitle{Investigating Perceived Message Credibility and Detection Accuracy of Fake and Real Information Across Information Types and Modalities}. In \bibinfo{booktitle}{\emph{Extended Abstracts of the 2023 CHI Conference on Human Factors in Computing Systems}} \emph{(\bibinfo{series}{CHI EA '23})}. \bibinfo{publisher}{Association for Computing Machinery}, \bibinfo{address}{New York, NY, USA}, \bibinfo{pages}{1--7}.
\newblock
\showISBNx{978-1-4503-9422-2}
\href{https://doi.org/10.1145/3544549.3585719}{doi:\nolinkurl{10.1145/3544549.3585719}}


\bibitem[Wittenberg et~al\mbox{.}(2024)]%
        {wittenberg_labeling_2024}
\bibfield{author}{\bibinfo{person}{Chloe Wittenberg}, \bibinfo{person}{Ziv Epstein}, \bibinfo{person}{Adam~J. Berinsky}, {and} \bibinfo{person}{David~G. Rand}.} \bibinfo{year}{2024}\natexlab{}.
\newblock \showarticletitle{Labeling {AI}-Generated Content: Promises, Perils, and Future Directions}.
\newblock \bibinfo{journal}{\emph{An {MIT} Exploration of Generative {AI}}} (\bibinfo{year}{2024}).
\newblock
\href{https://doi.org/10.21428/e4baedd9.0319e3a6}{doi:\nolinkurl{10.21428/e4baedd9.0319e3a6}}


\bibitem[Yaqub et~al\mbox{.}(2020)]%
        {yaqub_effects_2020}
\bibfield{author}{\bibinfo{person}{Waheeb Yaqub}, \bibinfo{person}{Otari Kakhidze}, \bibinfo{person}{Morgan~L. Brockman}, \bibinfo{person}{Nasir Memon}, {and} \bibinfo{person}{Sameer Patil}.} \bibinfo{year}{2020}\natexlab{}.
\newblock \showarticletitle{Effects of Credibility Indicators on Social Media News Sharing Intent}. In \bibinfo{booktitle}{\emph{Proceedings of the 2020 {CHI} Conference on Human Factors in Computing Systems}} \emph{(\bibinfo{series}{CHI '20})}. \bibinfo{publisher}{Association for Computing Machinery}, \bibinfo{address}{New York, NY, USA}, \bibinfo{pages}{1--14}.
\newblock
\showISBNx{978-1-4503-6708-0}
\href{https://doi.org/10.1145/3313831.3376213}{doi:\nolinkurl{10.1145/3313831.3376213}}


\bibitem[Yu et~al\mbox{.}(2024)]%
        {yu_fake_2024}
\bibfield{author}{\bibinfo{person}{Xiaomin Yu}, \bibinfo{person}{Yezhaohui Wang}, \bibinfo{person}{Yanfang Chen}, \bibinfo{person}{Zhen Tao}, \bibinfo{person}{Dinghao Xi}, \bibinfo{person}{Shichao Song}, \bibinfo{person}{Simin Niu}, {and} \bibinfo{person}{Zhiyu Li}.} \bibinfo{year}{2024}\natexlab{}.
\newblock \bibinfo{title}{Fake Artificial Intelligence Generated Contents ({FAIGC}): A Survey of Theories, Detection Methods, and Opportunities}.
\newblock
\href{https://doi.org/10.48550/arXiv.2405.00711}{doi:\nolinkurl{10.48550/arXiv.2405.00711}}
\showeprint[arxiv]{2405.00711}


\bibitem[Zhang et~al\mbox{.}(2024)]%
        {zhang2024decoding}
\bibfield{author}{\bibinfo{person}{Zhihui Zhang}, \bibinfo{person}{Josep~M Fort}, {and} \bibinfo{person}{Lluis Gim{\'e}nez~Mateu}.} \bibinfo{year}{2024}\natexlab{}.
\newblock \showarticletitle{Decoding emotional responses to AI-generated architectural imagery}.
\newblock \bibinfo{journal}{\emph{Frontiers in psychology}}  \bibinfo{volume}{15} (\bibinfo{year}{2024}), \bibinfo{pages}{1348083}.
\newblock


\end{thebibliography}

\appendix
\section{Survey Questions and Image Presented}
\label{appendix: survey questions}

\begin{figure*}[h]
    \centering
    \includegraphics[width=1.8\columnwidth]{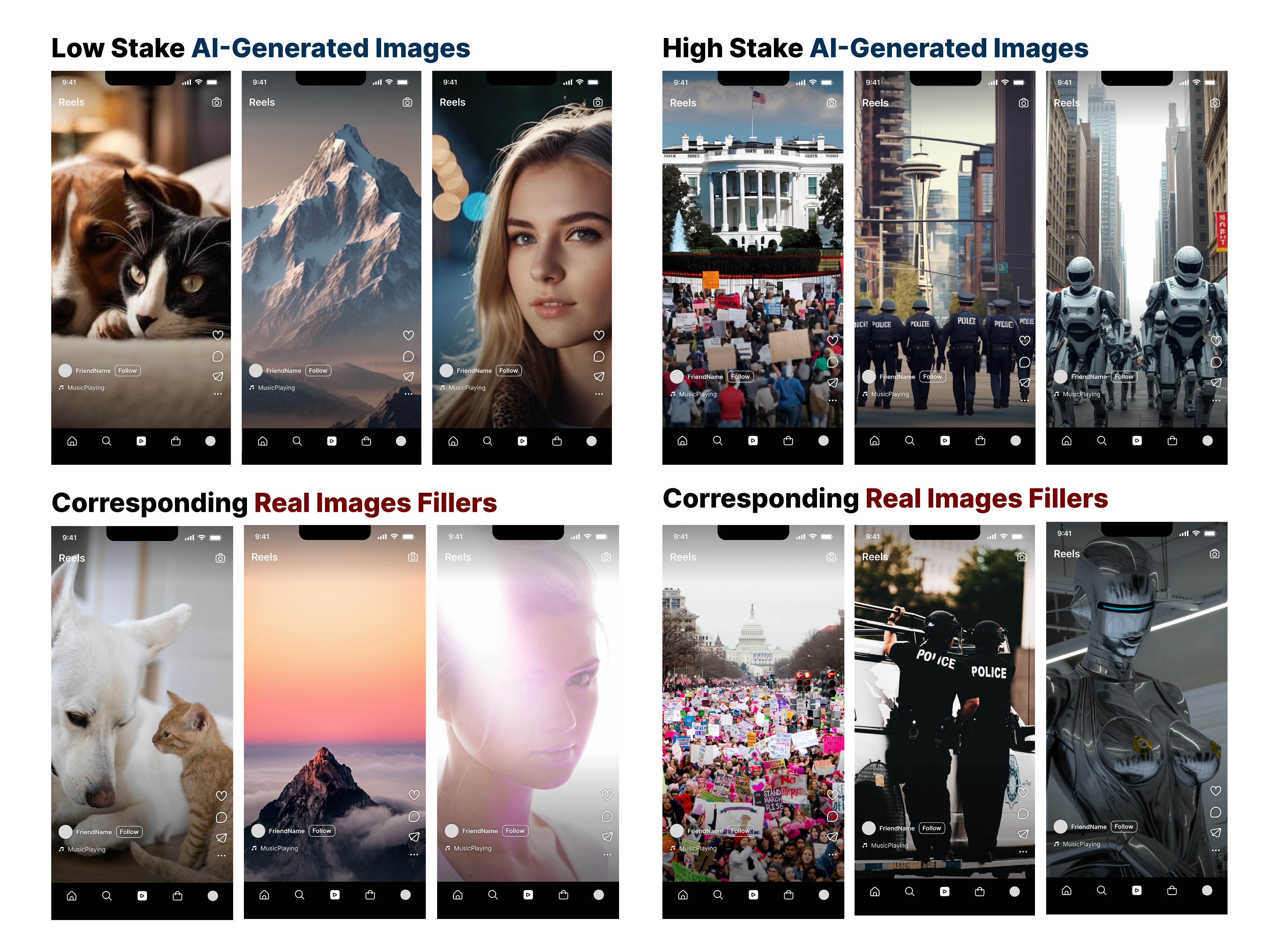}
    \caption{AI-generated and real image stimuli by content stakes. AI-generated images included low-stakes (pets, mountains, portraits) and high-stakes (protests, law enforcement, futuristic scenes); corresponding real-image fillers reduced priming and preserved ecological validity.}
    \label{fig: Image presented}
\end{figure*}

\textbf{Image Evaluation}
\begin{enumerate}[leftmargin=*]
  \item If you saw this content on a social media platform, how likely would you be to:
  \begin{itemize}
    \item Share this content
    \item Like or favorite this content
    \item Comment on or reply to this content
    \item Seek out additional information about this content’s topic
  \end{itemize}
  \FiveLikertLikely

  \item To what extent do you agree with the following statements about this content?
  \begin{itemize}
    \item This content is believable
    \item This content appears to be manipulated
  \end{itemize}
  \FiveLikert

  \item How much effort do you think it took to create this content? \ \SevenEffort

  \item Do you perceive the creation of this content to be mostly automated or requiring significant manual work? \ \SevenAutoManual

  \item In your opinion, to what extent is this \emph{AI-Generated} label and its information:
  \begin{itemize}
    \item Clear
    \item Informative
    \item Helpful
    \item Trustworthy
  \end{itemize}
  \FiveLikert
\end{enumerate}

\textbf{General Attitudes towards AI}
\begin{enumerate}[leftmargin=*]
  \item Do you think the increasing prevalence of AI-generated images on social media is a positive or negative development? \ (1: Very negative -- 7: Very positive)
  \item How familiar are you with the concept of AI and how it is used to create images? \ (1: Not familiar at all -- 7: Extremely familiar)
  \item Have you ever personally used an AI image generator? (Yes / No)
  \item If yes, approximately how many times have you used an AI image generator? (1--2 times, 3--5 times, 6--10 times, 10+ times)
  \item What are your primary purposes for using the AI image generator? (Artistic Creation, Entertainment, Educational Purposes, Professional Use, Other)
\end{enumerate}

\textbf{Social Media Usage}
\begin{enumerate}[leftmargin=*]
  \item Which social media platforms do you use at least 2--3 times a week? (Instagram, Facebook, Twitter, YouTube, TikTok, Other)
  \item How often do you notice or encounter images on social media that you believe were generated by AI? \ (1: Never -- 7: Very often)
  \item On which platforms do you encounter AI-generated images most frequently? (Carried-forward choices)
  \item In what types of social media content have you seen AI-generated images? (Short Videos, Image Posts, Advertisements, Profile Pictures, Other)
\end{enumerate}

\textbf{Demographics}
\begin{enumerate}[leftmargin=*]
  \item What is your age? (Under 18, 18--22, 23--27, 28--32, 33--37, 38--42, 43--47, 48--52, 52+)
  \item What is your gender identity? (Male, Female, Non-binary/Third gender, Prefer not to say, Other)
  \item What is the highest degree or level of school you have completed? (High School Diploma/GED, Bachelor’s Degree, Master’s Degree, Ph.D. or higher, Prefer not to say)
  \item What is your major/area? (Arts and Humanities, Business, Health and Medicine, STEM, Social Sciences, Other)
\end{enumerate}

\end{document}